\documentclass[aps,pre,twocolumn,showpacs,endfloats]{revtex4}
\usepackage{bm}
\begin{document}
\title{Force Equation That Describes the Quantum States of a Relativistic Spinless Particle}
\author{C. A. Ordonez}
\affiliation{Department of Physics, University of North Texas,  Denton, Texas 76203}
\date{\today}
\begin{abstract}
Newton's second law may be used to obtain a wave equation, which reduces to Schrodinger's equation in the nonrelativistic limit and for a conservative force.
\end{abstract}
\maketitle

\section{Introduction}

A quantum mechanical version of Newton's second law has existed for some time \cite{Bohm52A, Bohm52B}.  By substituting $\psi = \sqrt{f } e^{i S/\hbar } $ into Schrodinger's equation, it is possible to arrive at a continuity equation, 
\begin{equation}
{\partial  f \over \partial t} + {\bm \nabla } \cdot \left( f \bm{v} \right) = 0 ,
\end{equation}
and the quantum mechanical version of Newton's second law, 
\begin{equation}
m {d^2{\bm r} \over dt^2 } = - {\bm \nabla } \left(  V +  Q \right) ,
\end{equation}
provided $\bm{v} = {\bm \nabla } S  / m $ is velocity, and 
\begin{equation}
 Q = {\hbar^2 \over 2 m}  \left[ {({\bm \nabla } f) \cdot ({\bm \nabla } f) \over (2 f)^2 }   - \left( { \nabla^2 f  \over 2 f} \right) \right]  
\end{equation}
is a quantum mechanical potential energy \cite{Bohm52A, Bohm52B}.  Here, 
$\psi $ is the wavefunction,  $f$ is the probability density, $V$ is the classical potential energy, $\bm{r} $ is the configuration space coordinate vector, $t$ is time, $m$ is the particle's mass, $i = \sqrt{-1} $, $\hbar = h / (2 \pi )$, and $h$ is Planck's constant.
This type of formulation has been used as the basis of a significant number of recent studies \cite{Holland05,Rassolov06}.
Although Newton's second law describes the trajectory of a point particle, Newton's second law may also be used to obtain a wave equation that describes the quantum states of a relativistic spinless particle.  

\section{Preliminaries}

As a starting point, it is assumed that a time-dependent probability density in configuration space, $f (\bm{r} , t ) $, and a time-dependent probability density in momentum space, $g_p(\bm{p} , t) $, describe the state of a spinless particle.  A wavevector $\bm{k} $ is defined in relation to momentum $\bm{p} $ by $\bm{p} = \hbar \bm{k} $.  Wavefunctions in configuration space, momentum space, and wavevector space, denoted $\psi (\bm{r} , t ) $,  $\chi_p (\bm{p} , t )$, and  $\chi_k (\bm{k} , t )$, respectively, are defined in relation to probability densities in configuration space, momentum space, and wavevector space, denoted $f (\bm{r} , t ) $,  $g_p (\bm{p} , t )$, and $g_k (\bm{k} , t )$, respectively, such that $ f(\bm{r} , t) = \psi^* (\bm{r} , t) \psi (\bm{r} , t ) $, $ g_p (\bm{p} , t) = \chi_p^* (\bm{p} , t) \chi_p (\bm{p} , t) $, and $ g_k (\bm{k} , t) = \chi_k^* (\bm{k} , t) \chi_k (\bm{k} , t) $.  Here, a complex conjugate is indicated with a superscript star $(^*)$.  As a postulate, $\chi_k (\bm{k} , t )$ is taken to be the three-dimensional Fourier transform of $\psi (\bm{r} , t ) $,
\begin{equation}
\label{FourierTransform}
\chi_k (\bm{k} , t ) = {1 \over (2 \pi )^{3/2} } \int \psi (\bm{r} , t ) e^{- i \bm{k} \cdot \bm{r} } d^3 r ,
\end{equation} 
where $\int  d^3 r $ is an integration over configuration space.  The inverse transform is 
\begin{equation}
\label{InverseFourierTransform}
\psi (\bm{r} , t ) = {1 \over (2 \pi )^{3/2} } \int \chi_k (\bm{k} , t ) e^{ i \bm{k} \cdot \bm{r} } d^3 k ,
\end{equation} 
where $\int  d^3 k $ is an integration over wavevector space.

The expectation value of an arbitrary expression $A(\bm{p} )$ that is written in terms of momentum is evaluated in momentum space as
\begin{equation}  
\label{Ap}
\left< A(\bm{p} ) \right>  = \int A(\bm{p} ) g_p (\bm{p} , t ) d^3 p = \int A(\bm{p} ) \chi_p^* (\bm{p} , t ) \chi_p (\bm{p} ,t ) d^3 p ,
\end{equation} 
where $\int  d^3 p $ is an integration over momentum space.
The same expectation value can be evaluated as an integration over wavevector space using
\begin{equation}  
\label{Ak}
\left< A(\bm{p} ) \right>  =  \int A(\hbar \bm{k} ) \chi_k^* (\bm{k} , t) \chi_k (\bm{k} ,t) d^3 k .
\end{equation} 
The expectation value can be evaluated as an integration over configuration space using
\begin{equation}  
\label{ArO}
\left< A(\bm{p} ) \right>  =  \int \psi^* (\bm{r} ,t )  O [  \psi (\bm{r} ,t ) ]  d^3 r  ,
\end{equation} 
where $O$ is a differential operator in configuration space, and 
$O ( \psi )$ is understood to indicate that $O$ operates on $\psi $.
Substituting $\psi $ given by Eq.~(\ref{InverseFourierTransform}) into Eq.~(\ref{ArO}) gives
\begin{equation}  
\label{Ar2}
\left< A(\bm{p} ) \right>  =  {1 \over (2 \pi )^{3} }  \int 
\chi_k^* (\bm{k}_1 , t ) 
\chi_k (\bm{k} , t ) 
e^{ - i \bm{k}_1 \cdot \bm{r} }
O \left( e^{ i \bm{k} \cdot \bm{r} } \right) 
d^3 r d^3 k_1 d^3 k
\end{equation} 
where the subscript on $\bm{k}_1 $ is used to distinguish two different sets of wavevector space integration variables.  
An expression for $O$ that allows the expectation value to be evaluated using Eq.~(\ref{ArO}) is one that satisfies  
\begin{equation}  
\label{O}
O \left( e^{ i \bm{k} \cdot \bm{r} }  \right)  =   A(\hbar \bm{k} ) e^{ i \bm{k} \cdot \bm{r} } .
\end{equation} 
Substituting Eq.~(\ref{O}) into Eq.~(\ref{Ar2}) yields 
\begin{equation}  
\label{Ar3}
\left< A(\bm{p} ) \right>  = 
\int A(\hbar \bm{k} ) \chi_k^* (\bm{k}_1 , t ) \chi_k (\bm{k} , t )    \delta (\bm{k} - \bm{k}_1 ) d^3 k_1 d^3 k ,
\end{equation} 
where $\delta (\bm{k} - \bm{k}_1 ) $ is the Dirac delta function, as represented by 
\begin{equation}  
\label{delta}
\delta (\bm{k} - \bm{k}_1 ) = {1 \over (2 \pi )^3 } \int e^{ i (\bm{k} - \bm{k}_1 ) \cdot \bm{r} } d^3 r  .
\end{equation} 
Carrying out the integration $\int d^3 k_1$ in Eq.~(\ref{Ar3}) yields Eq.~(\ref{Ak}).

\section{Differential Expressions}

Note that Eq.~(\ref{ArO}) can be written as
\begin{equation}  
\label{ArD}
\left< A(\bm{p} ) \right>  =  \int A(\bm{p} )  f (\bm{r} ,t )   d^3 r  ,
\end{equation} 
where $A(\bm{p} )$ in configuration space is understood to represent a differential expression defined by
\begin{equation}  
\label{D}
A(\bm{p} )   =  {O [  \psi (\bm{r} ,t ) ] \over \psi (\bm{r} ,t ) } .
\end{equation} 
For example, the expectation value of $\bm{p} $ would be written as $\left< \bm{p} \right> = \int \bm{p} f (\bm{r} , t ) d^3 r$, where $\bm{p} $ represents a differential expression in configuration space.  Two differential expressions that are defined according to Eq.~(\ref{D}) from operators that
satisfy Eq.~(\ref{O}) are
\begin{equation}  
\label{momentum}
\bm{p} = {- i \hbar  \bm{\nabla } \psi (\bm{r} ,t ) \over \psi (\bm{r} ,t ) }  ,
\end{equation} 
and
\begin{equation}
p^2 =   {- \hbar^2 \nabla^2 \psi (\bm{r} ,t ) \over \psi (\bm{r} ,t ) }  .
\end{equation} 
It can be shown that $p^2 = \bm{p} \cdot \bm{p}^* $ when $\psi = e^{ i \bm{k} \cdot \bm{r} } $.

\section{Relativistic Equation}

The time rate of change of the differential expression, $ \bm{p} (\bm{r} , t ) $, is
\begin{equation}
{d\bm{p} \over dt} =  {\partial \bm{p} \over \partial t}  + \left( \bm{v} \cdot \bm{\nabla } \right) \bm{p} = {\partial \bm{p} \over \partial t}  +  {\left( \bm{p} \cdot \bm{\nabla } \right) \bm{p} \over m_\gamma } . 
\end{equation}
Here, velocity is written in terms of relativistic momentum as $\bm{v} = \bm{p} / m_\gamma $, where
\begin{equation}  
\label{mgamma}
m_\gamma = m \sqrt{ 1 + {p^2 \over (m c)^2 } }  ,
\end{equation} 
$m$ is the particle's rest mass, and $c$ is the speed of light.  The expression $\bm{v} = \bm{p} / m_\gamma $ is obtained by inverting  $\bm{p} = m \bm{v} / \sqrt{1-v^2/c^2}$.  According to Newton's second law, $\bm{F} = d \bm{ p } /dt  $, where $ \bm{F} $ is the sum of the forces that act on the particle, and $d \bm{ p } /dt   $ is the time rate of change of the particle's relativistic momentum.  With Newton's second law, and employing the vector identity, 
$\bm{\nabla } {(\bm{a} \cdot \bm{b} )} = \bm{a} \times (\bm{\nabla } \times \bm{b} ) + \bm{b} \times (\bm{\nabla } \times \bm{a} ) + (\bm{a} \cdot \bm{\nabla } ) \bm{b}  + (\bm{b} \cdot \bm{\nabla } ) \bm{a} $, where $\bm{a}$ and $\bm{b}$ are vectors, a force equation that describes the quantum states of a relativistic spinless particle is written as 
\begin{equation}
\label{QFrelativistic} 
\bm{F}  = {\partial \bm{p} \over \partial t}  +  {\bm{\nabla } p^2 \over 2 m_\gamma } - {\bm{p} \times \left( \bm{\nabla } \times  \bm{p}  \right) \over m_\gamma } .
\end{equation}
Equation (\ref{QFrelativistic}) may be considered a wave equation, because $\bm{p}$ and $p^2$ represent differential expressions that contain the wavefunction $\psi $.  Equation (\ref{QFrelativistic}) applies even when the presence of a nonconservative force does not allow a potential energy to be defined.

\section{Nonrelativistic, Conservative-Force Equation}

In the nonrelativistic $(m_\gamma \rightarrow m)$ limit, and for a conservative force (${\bm F}   = -  {\bm \nabla }  V$, where $V$ is potential energy), Eq.~(\ref{QFrelativistic}) is written as
\begin{equation}
\label{QFnonrelativistic} 
-  {\bm \nabla }  V  = {\partial \bm{p} \over \partial t} + \bm{\nabla } \left( {p^2 \over 2 m } \right) - {\bm{p} \times \left( \bm{\nabla } \times  \bm{p}  \right) \over m } .
\end{equation} 
The first term on the right is written as
\begin{equation}
\label{firstterm}
{\partial \bm{p} \over \partial t} = {\partial \over \partial t}  \left( - i \hbar   {\bm{\nabla } \psi  \over \psi } \right) =  \bm{\nabla } \left( {- i \hbar \over \psi }  {\partial  \psi \over \partial t}  \right)  ,
\end{equation}
where the second equality is arrived at by direct substitution of $ \psi (\bm{r} ,t )$.
The second term on the right in Eq.~(\ref{QFnonrelativistic}) represents the gradient of the classical kinetic energy.  The associated differential expression is
\begin{equation}
\label{kineticenergy}
\bm{\nabla } \left( {p^2 \over 2 m } \right) =  \bm{\nabla } \left( {- \hbar^2 \over 2 m}  {\nabla^2 \psi \over \psi  }  \right) .
 \end{equation}
The third term on the right in Eq.~(\ref{QFnonrelativistic}) is zero, 
\begin{equation}
\label{thirdterm}
\bm{p} \times \left( \bm{\nabla } \times  \bm{p}  \right) = 0 .
 \end{equation}
 
With Eqs.~(\ref{firstterm}) - (\ref{thirdterm}), Eq.~(\ref{QFnonrelativistic}) is written as
\begin{equation}
\label{TDSE0} 
 \bm{\nabla } \left( { [- \hbar^2 / (2 m)]  \nabla^2 \psi + V \psi -  i \hbar (\partial \psi / \partial t ) \over \psi } \right) = 0 . 
\end{equation}
For $\psi = \psi (\bm{r} , t)$, Eq.~(\ref{TDSE0}) is satisfied if the numerator of the quotient equals zero. Setting the numerator equal to zero, the resulting equation can be written as Schrodinger's equation,
\begin{equation}
\label{TDSE} 
 - {\hbar^2 \over  2 m } \nabla^2 \psi +  V \psi =  i \hbar {\partial \psi  \over \partial t }  .
\end{equation}
For a stationary state, Eq.~(\ref{QFnonrelativistic}) is written as $ \bm{\nabla }  [p^2 / (2 m) +  V] = 0$, where ${\partial \bm{p} \over \partial t}  = 0$ and Eq.~(\ref{thirdterm}) are used.  The expression $ {\bm \nabla } [p^2 / (2 m) +  V] =  0 $ is satisfied by a spatially constant classical energy $E  = p^2 / (2 m) +  V $.  Upon substituting $\psi $ into $E  = p^2 / (2 m) +  V  $, the resulting equation can be written as the time-independent version of Schrodinger's equation,
\begin{equation}
\label{TISE} 
 - {\hbar^2 \over  2 m } \nabla^2 \psi +  V \psi = E \psi .
\end{equation}

\section{Relativistic, Conservative-Force Equation}

Relativistic versions of Eqs.~(\ref{TDSE}) and (\ref{TISE}) are readily obtained, provided the
second term on the right in Eq.~(\ref{QFrelativistic}) is rewritten as the gradient of the relativistic energy:
\begin{equation}
\label{relativisticenergy}
 { \bm{\nabla } p^2 \over 2 m_\gamma } = \bm{\nabla } \left[ \sqrt{ (m c^2)^2 + c^2 p^2 } \right] .
\end{equation}
The associated differential expression is
\begin{equation}
\label{secondterm}
 { \bm{\nabla } p^2 \over 2 m_\gamma } =\bm{\nabla } \left[ \sqrt{ (m c^2 )^2 - (\hbar c)^2 {\nabla^2 \psi \over \psi } } \right] .
\end{equation}
With ${\bm F} = -  {\bm \nabla }  V$, and with Eqs.~(\ref{firstterm}), (\ref{thirdterm}), and (\ref{secondterm}), Eq.~(\ref{QFrelativistic}) is written as
\begin{equation}
\label{QFvectorrel} 
 \bm{\nabla } \left( { \sqrt{ (m c^2 )^2 - (\hbar c)^2 {(\nabla^2 \psi ) / \psi \, } } \, \psi + V \psi -  i \hbar (\partial \psi / \partial t ) \over \psi } \right)
  = 0  .
\end{equation}
With $\psi = \psi (\bm{r} , t)$, the numerator of the quotient is set equal to zero, and the resulting equation is written as
\begin{equation}
\label{TDRE}
\sqrt{ (m c^2 )^2 - (\hbar c)^2 {\nabla^2 \psi \over \psi } } \, \psi + V \psi =  i \hbar {\partial \psi \over \partial t }  .
\end{equation}
The stationary state version of Eq.~(\ref{TDRE}) is
\begin{equation}
\sqrt{ (m c^2 )^2 - (\hbar c)^2 {\nabla^2 \psi \over \psi } } \, \psi + V \psi = E \psi ,
\end{equation}
where $E$ now represents the sum of the relativistic energy and the potential energy.

\section{Conclusion}

In summary, a way to arrive at Schrodinger's equation from Newton's second law was developed by writing momentum expressions as differential expressions.  Newton's second law was used to write a force  equation, Eq.~(\ref{QFrelativistic}), which represents a wave equation that describes the quantum states of a relativistic spinless particle.  Equation (\ref{QFrelativistic}) was used to obtain Schrodinger's equation (both time-dependent and time-independent versions) in the nonrelativistic limit and for a conservative force.  Relativistic versions of both versions of Schrodinger's equation were obtained.  

The author would like to thank Profs.~J.~Kowalski, D.~Kobe, P.~ Grigolini, and S.~Quintanilla  for comments and suggestions.  This material is based upon work supported by the Department of Energy under Grant No.~DE-FG02-06ER54883.


\begin{thebibliography}{}
\bibitem{Bohm52A} D. Bohm, Phys. Rev. {\bf 85}, 166 (1952).
\bibitem{Bohm52B} D. Bohm, Phys. Rev. {\bf 85}, 180 (1952).
\bibitem{Holland05} P. Holland, Annals of Physics {\bf 315}, 505 (2005); and references therein.
\bibitem{Rassolov06} V. A. Rassolov,  S. Garashchuk, and G. C. Schatz, J. Phys. Chem. A {\bf 110}, 5530 (2006); and references therein.
\end{thebibliography}
\end{document}